\documentstyle[12pt,psfig]{article}
\catcode`\@=11
\def\marginnote#1{}
\newcount\hour
\newcount\minute
\newtoks\amorpm
\hour=\time\divide\hour by60
\minute=\time{\multiply\hour by60 \global\advance\minute by-
\hour}
\edef\standardtime{{\ifnum\hour<12 \global\amorpm={am}%
    \else\global\amorpm={pm}\advance\hour by-12 \fi
    \ifnum\hour=0 \hour=12 \fi
    \number\hour:\ifnum\minute<100\fi\number\minute\the\amorpm}}
\edef\militarytime{\number\hour:\ifnum\minute<100\fi\number\minute}
\def\draftlabel#1{{\@bsphack\if@filesw {\let\thepage\relax
  \xdef\@gtempa{\write\@auxout{\string
    \newlabel{#1}{{\@currentlabel}{\thepage}}}}}\@gtempa
    \if@nobreak \ifvmode\nobreak\fi\fi\fi\@esphack}
     \gdef\@eqnlabel{#1}}
\def\@eqnlabel{}
\def\@vacuum{}
\def\draftmarginnote#1{\marginpar{\raggedright\scriptsize\tt#1}}
\def\draft{\oddsidemargin -.5truein
        \def\@oddfoot{\sl preliminary draft \hfil
        \rm\thepage\hfil\sl\today\quad\militarytime}
        \let\@evenfoot\@oddfoot \overfullrule 3pt
        \let\label=\draftlabel
        \let\marginnote=\draftmarginnote

\def\@eqnnum{(\theequation)\rlap{\kern\marginparsep\tt\@eqnlabel}%
\global\let\@eqnlabel\@vacuum}  }
\def\preprint{\twocolumn\sloppy\flushbottom\parindent 1em
        \leftmargini 2em\leftmarginv .5em\leftmarginvi .5em
        \oddsidemargin -.5in    \evensidemargin -.5in
        \columnsep 15mm \footheight 0pt
        \textwidth 250mmin      \topmargin  -.4in
        \headheight 12pt \topskip .4in
        \textheight 175mm
        \footskip 0pt

\def\@oddhead{\thepage\hfil\addtocounter{page}{1}\thepage}
        \let\@evenhead\@oddhead \def\@oddfoot{} \def\@evenfoot{}
}
\def\titlepage{\@restonecolfalse\if@twocolumn\@restonecoltrue\onecolumn
     \else \newpage \fi \thispagestyle{empty}\c@page\z@
        \def\thefootnote{\fnsymbol{footnote}} }
\def\endtitlepage{\if@restonecol\twocolumn \else  \fi
        \def\thefootnote{\arabic{footnote}}
        \setcounter{footnote}{0}}  %\c@footnote\z@ }
\catcode`@=12
\relax
\def\be{\begin{equation}}
\def\ee{\end{equation}}
\def\bea{\begin{eqnarray}}
\def\eea{\end{eqnarray}}
\def\simlt{\stackrel{<}{{}_\sim}}
\def\simgt{\stackrel{>}{{}_\sim}}

\def\NPB#1#2#3{{\it Nucl.~Phys.} {\bf{B#1}} (19#2) #3}
\def\PLB#1#2#3{{\it Phys.~Lett.} {\bf{B#1}} (19#2) #3}
\def\PRD#1#2#3{{\it Phys.~Rev.} {\bf{D#1}} (19#2) #3}
\def\PRL#1#2#3{{\it Phys.~Rev.~Lett.} {\bf{#1}} (19#2) #3}

\def\PTP#1#2#3{{\it Prog.~Theor.~Phys.} {\bf#1}  (19#2) #3}

\relax
\begin{document}
\topmargin-2.5cm
%\draft
%\preprint
%
\begin{titlepage}
\begin{flushright}
CERN-TH/95-45\\
DESY 95-038 \\
IEM-FT-103/95 \\
hep--ph/9504316 \\
\end{flushright}
\vskip 0.3in
\begin{center}{\Large\bf
Analytical expressions for radiatively corrected  \\
Higgs masses and couplings in the MSSM
\footnote{Work supported in part by
the European Union (contract CHRX-CT92-0004) and
CICYT of Spain
(contract AEN94-0928).}  } \\
\vskip .5in
{\bf M. Carena}$^{\ddag}$,
{\bf J.R. Espinosa}$^{\S}$,
{\bf M. Quir\'os}$^{\ddag}$
\footnote{On leave of absence from Instituto
de Estructura de la Materia, CSIC, Serrano 123, 28006-Madrid,
Spain.}
and {\bf C.E.M. Wagner}$^{\ddag}$ \vskip.5in
$^{\ddag}$CERN, TH Division, CH--1211 Geneva 23, Switzerland\\
$^{\S}$Deutsches Elektronen-Synchrotron DESY, Hamburg, Germany \\
\end{center}
\vskip.5cm
\begin{center}
{\bf Abstract}
\end{center}
\begin{quote}
We propose, for the computation of the Higgs mass spectrum
and couplings, a renormalization-group improved
leading-log approximation, where the renormalization
scale is fixed to the top-quark pole mass.
For the case $ m_A\sim M_{\rm SUSY}$,
our leading-log approximation differs
by less than 2 GeV from previous results on the Higgs mass
computed using a nearly scale independent
renormalization-group
improved effective potential up to next-to-leading order.
Moreover, for the general case $m_A\simlt M_{\rm SUSY}$,
we provide analytical formulae
(including two-loop leading-log corrections)
for all the masses and couplings in
the Higgs sector. For $M_{\rm SUSY}\simlt 1.5$ TeV and
arbitrary values of $m_A$, $\tan\beta$ and the stop mixing
parameters, they reproduce the numerical renormalization-group
improved leading-log
result for the Higgs masses with an error of less than 3 GeV.
For the Higgs couplings, our analytical formulae
reproduce the numerical results equally well. Comparison with other
methods is also performed.

\end{quote}
\vskip1.cm

\begin{flushleft}
CERN-TH/95-45\\
March 1995 \\
\end{flushleft}

\end{titlepage}
\setcounter{footnote}{0}
\setcounter{page}{0}
\newpage
%
% BODY
\noindent
\section{Introduction}

The most appealing extension of the Standard Model (SM) is the Minimal
Supersymmetric Standard Model (MSSM), which can provide a technical
explanation for the hierarchy stability from $M_{\rm Pl}$ to the electroweak
scale. From the experimental point of view, the MSSM is also attractive and
presents clear and distinct signatures. In particular, its Higgs sector
predicts the existence of a light CP-even state with a mass which cannot
exceed a value of $\sim$ 150 GeV, unlike the case of the SM Higgs whose
mass can reach much larger values ($\sim$ 1 TeV). The precise bounds on the
mass spectrum of the MSSM Higgs bosons strongly
depend on the value of the top-quark
mass $M_t$~\cite{topc,ERZ}.

In view of Higgs searches at future colliders, and of
the fact that the top-quark mass is being measured at the
Tevatron~\cite{CDF,D0},
it is of the highest interest to provide determinations as precise as possible
of the
radiatively corrected Higgs mass spectrum and Higgs couplings and, in
particular,
to find accurate upper bounds on the mass of the lightest Higgs boson.
To this end one can assume that $M_W\ll M_{\rm SUSY}\simlt
 {\cal O}$(few) TeV, where $M_{\rm SUSY}$ is the characteristic
supersymmetry particle mass scale  and its
upper limit is determined by naturalness arguments.
In this case the upper bound on the lightest Higgs mass
is reached when the mass of the CP-odd Higgs,
$m_A$, is of the order of $M_{\rm SUSY} $, and the theory below
$M_{\rm SUSY} $ is the SM with threshold effects
at $M_{\rm SUSY} $. For values of $m_A<
M_{\rm SUSY} $ the theory below $M_{\rm SUSY}$
is the two-Higgs doublet model, and then, two CP-even and
one CP-odd neutral Higgs bosons,
and two charged Higgs bosons appear in the physical spectrum below
$M_{\rm SUSY} $.

It is the aim of this letter to explore the whole range
$m_A\simlt M_{\rm SUSY} $
and provide an accurate numerical evaluation as well as
analytical formulae for the whole Higgs mass spectrum and Higgs
couplings of the MSSM. We shall also compute the range of validity
of our approximations, as well as make comparison with other
methods currently used in the literature.

\section{The case $m_A\sim M_{\rm SUSY}$}

This case is well described, as we mentioned above, by the SM with
the quartic coupling
\be
\label{quartic}
\lambda=\frac{1}{4}(g_2^2+g_1^2)\cos^22\beta
\ee
at $ M_{\rm SUSY} $, and one-loop threshold contributions at that scale
\be
\label{threshold}
\Delta\lambda=\frac{3}{16\pi^2} \left(h_t^{\rm SM}\right)^4
\ \tilde{X}_t.
\ee
In the above, $g_2$ and $g_1$ are the $SU(2)\times U(1)_Y$ gauge couplings,
$\tan\beta=v_2/v_1$ is the ratio of the vacuum expectation values of
the neutral components of the two supersymmetric Higgs bosons, $h_t^{\rm SM}$
is the SM top-quark Yukawa coupling and
$\tilde{X}_t$ is the stop mixing parameter~\footnote{In
this section we are neglecting
the sbottom mixing parameter.}:
\begin{eqnarray}
\label{stopmix}
\tilde{X}_{t} & =& \frac{2 \tilde{A}_t^2}{M_{\rm SUSY}^2}
                  \left(1 - \frac{\tilde{A}_t^2}{12 M_{\rm SUSY}^2} \right)
\nonumber \\
\tilde{A}_t & = & A_t-\mu\cot\beta.
\end{eqnarray}
The supersymmetric scale
 $M_{\rm SUSY}$ is taken as the squared root of the arithmetical average of
the diagonal entries in the squared squark mass matrix and,
for simplicity, we assume that all supersymmetric particle masses
are of order $M_{\rm SUSY}$. A more detailed discussion about the
definition of $M_{\rm SUSY}$ and the possible effect of light sparticles
will be given in sections 4 and 5.

The Higgs boson mass was first determined by
renormalization-group resummation
of all-loop leading-log (LL) corrections~\cite{LL,CSW,CPR,HH}.
Some next-to-leading-log
(NTLL) corrections were further introduced~\cite{EQ1,HHo}
and finally a complete NTLL analysis was
performed in Refs.~\cite{KYS,CEQR}.

In Ref.~\cite{CEQR} physical pole masses for the top-quark
and Higgs-boson were used and the dependence with respect to the
renormalization-group scale  in the NTLL approximation was explicitly
analysed.
One of the main issues in~\cite{CEQR} was the comparison between the LL
and NTLL approximations. As expected, the LL approximation shows strong
scale dependence, while the NTLL is almost scale independent. This implies
not only that the choice of the renormalization-group scale is irrelevant when
working in the NTLL approximation, but also that a judicious choice of the
renormalization scale in the LL approximation can yield very accurate
results. It turns out that, independently of the value of the stop
mixing parameter,  the scale where both results coincide  is very
close to the pole top-quark mass $M_t$ (see e.g. Fig.~8 of Ref.~\cite{CEQR}).

This result can also be understood from Ref.~\cite{SZ} where the physical
Higgs mass is related to the
quartic coupling in the $\overline{\rm MS}$ scheme,
$\lambda(t)$. One can write
\be
\label{sirlin}
M_h^2=m_h^2(t)(1-\delta-\Delta r)
\ee
where
\be
\label{runmas}
m_h^2(t)=2\lambda(t)v^2
\ee
is the running Higgs mass, and
$v=174.1$ GeV. From~\cite{SZ,MS}, and neglecting
scalar and electroweak gauge couplings, we can write
\begin{eqnarray}
\delta+\Delta r & = & \frac{3}{16\pi^2}\left(h_t^{\rm SM}\right)^2
\left(2-4\frac{M_t^2}{M_h^2}\right)\log\frac{Q^2}{M_t^2}
+\varepsilon \nonumber \\ \label{deltas}
\varepsilon & = & -\frac{3}{16\pi^2} \left(h_t^{\rm SM}\right)^2
\left[Z\left(\frac{M_t^2}{M_h^2}\right)-2\right]\left(1-4
\frac{M_t^2}{M_h^2}\right)
\end{eqnarray}
where $Q$ is the renormalization scale and
the function $Z(x)$ is given, for $x>1/4$, by
\be
\label{Z}
Z(x)=2\sqrt{4x-1}\tan^{-1}(1/\sqrt{4x-1})\ .
\ee
We can see from (\ref{deltas}) that choosing $Q=M_t$, the difference
between the pole mass and the running mass is determined by
$\varepsilon$
($\varepsilon \leq $ 1.7\% for $M_h\geq 60$ GeV and $M_t\leq 200$~GeV).
For instance for $M_h=90$ GeV and $M_t=180$ GeV, then $\varepsilon
=1.3$\%. Other values of $Q$ (e.g. $Q=M_Z$ as can be usually found
in the literature~\cite{LP})
 can provide much larger corrections, depending on the
values of $M_h$ and $M_t$, unless an appropriate decoupling of the
top-quark effects below the top-quark mass scale is performed.

For the Higgs mass, we shall therefore adopt the value provided by
(\ref{runmas}) with $Q=M_t$ and work in the LL approximation as suggested
by (\ref{deltas}). We compare, in Fig.~1, the all-loop RG improved
NTLL~\cite{CEQR} (solid) and LL (dashed)
lines for $ M_{\rm SUSY} =1$ TeV. We consider
the case of zero mixing, $\tilde{X}_t = 0$, and the case
$\tilde{X}_t$ = 6, which maximizes the lightest CP-even Higgs mass,
large values of $\tan\beta$ ($\tan\beta=15$) and small values of $\tan\beta$
[the infrared (IR) fixed point solution with $\sin\beta\sim (200\ {\rm GeV})
/M_t$]. The RG improved LL aproximation involves the one--loop RG running
of the couplings, the one-loop threshold effects $\Delta\lambda$,
Eq.~(\ref{threshold}), from the decoupling of the supersymmetric particles at
$ M_{\rm SUSY}$, and the computation of the Higgs masses at the scale $Q=M_t$.
Moreover, in both approximations the pole top-quark mass
$M_t$ was computed from the on-shell running mass
$m_t$ through the corresponding one-loop QCD correction factor
\be
m_t=\frac{M_t}{1+\frac{4}{3\pi}\alpha_3(M_t)}.
\label{mtMt}
\ee
\
We see that, for all values of the top-quark mass, $\tan\beta$
and the stop mixing parameter,
the solid and dashed curves agree with each other with
an accuracy better than 2 GeV.

We have also worked out an analytical approximation to the numerical
all-loop renormalization-group improved LL result,
including two-loop
leading-log effects. It is given by
\begin{eqnarray}
m_h^2& = & M_Z^2\cos^2 2\beta\left( 1-\frac{3}{8\pi^2}\frac{m_t^2}
{v^2}\ t\right) \nonumber \\
\label{mhsm}
& + & \frac{3}{4\pi^2}\frac{m_t^4}{v^2}\left[ \frac{1}{2}\tilde{X}_t + t
+\frac{1}{16\pi^2}\left(\frac{3}{2}\frac{m_t^2}{v^2}-32\pi\alpha_3
\right)\left(\tilde{X}_t t+t^2\right) \right]
\end{eqnarray}
where the angle $\beta$ is defined here at the scale $M_{\rm SUSY}$,
\be
\label{escala}
t=\log\frac{ M_{\rm SUSY}^2}{M_t^2},
\ee
and
\be
\label{alfa3sm}
\alpha_3(M_t)=\frac{\alpha_3(M_Z)}{1+\frac{b_3}{4\pi}\alpha_3(M_Z)
\log(M_t^2/M_Z^2)},
\ee
where $b_3 = 11 - 2 N_f/3$ is the one-loop QCD beta function and
$N_f$ is the number of quark flavours, which is equal to 5
at scales below $M_t$.  Notice that Eq.~(\ref{mhsm}) includes
the leading $D$-term correction ${\cal O}(M_Z^2 m_t^2)$.
As was pointed out in Ref.~\cite{B}, this contribution cannot be neglected,
since, for the experimental $M_t$ range, it can account for a (negative)
shift as large as $\sim$ 5 GeV.
Formula (\ref{mhsm}) is also plotted in Fig.~1 (dotted lines), for the
same values of the supersymmetric parameters as before.
We can see that the
analytical approximation reproduces the numerical results within an
error of less than 2 GeV in all cases.

Finally, in Fig.~2 we have plotted the Higgs mass, in the three different
approximations, for $M_t=175$ GeV, as a function of
$M_{\rm SUSY}$. As expected the
NTLL (solid) and LL (dashed) curves coincide within less than 2 GeV
difference. The analytical approximation also remains within this
level of accuracy  if
$M_{\rm SUSY}\simlt 1.5$
TeV. Hence, it follows that for higher values of $M_{\rm SUSY}$,
agreement with numerical results would
demand the addition of ${\cal O}(t^3)$ terms (three-loop leading-log), but
for the theoretically preferred values of the supersymmetry breaking
scale, $M_{\rm SUSY}\simlt 1$ TeV, the analytical formula
provides excellent results.
For values of $M_{\rm SUSY} \leq$ 500 GeV, the maximal value of the mixing
parameter $\tilde{X}_t$ is restricted (see discussion in section 4).

\section{The case $m_A\simlt M_{\rm SUSY}$}

In the more general case, $m_A\simlt M_{\rm SUSY}$, the
effective theory below the scale $M_{\rm SUSY}$~\cite{HH} is a two-Higgs
doublet model where the tree level quartic couplings can be written in terms
of dimensionless parameters $\lambda_i$, $i=1,\dots ,\ 7$, whose tree level
values are functions of the gauge couplings, and with one-loop
threshold corrections $\Delta\lambda_i$, $i=1,\dots ,\ 7$, expressed as
functions of the supersymmetric Higgs mass $\mu$ and the soft supersymmetry
breaking parameters $A_t$, $A_b$ and $m_A$.
We are using the conventions of Ref.~\cite{HH}.
Motivated by the previous agreement among the  NTLL and LL approximations and
the analytical formulae,
we can safely extrapolate the analytical approach,
to obtain the two-loop results
for the Higgs spectrum within the MSSM.

The CP-even light and heavy Higgs masses
and the charged Higgs mass are given as functions of
$\tan \beta$, $M_{\rm SUSY}$, $A_t$, $A_b$, $\mu$, the CP-odd Higgs mass $m_A$
and the physical top-quark mass $M_t$ related to the on-shell running
mass $m_t$ through Eq.~(\ref{mtMt}).
We present below the masses and mixing angle of the Higgs
sector as functions of the parameters $\lambda_i$,
for which we  have derived the corresponding analytical
formulae and which  are the equivalent in the two-Higgs
doublet model of the  analytical approximation (\ref{mhsm}) for
the case of the SM. For completeness, we include the bottom
Yukawa coupling effects, which may become large for values of
$\tan\beta \simeq m_t/m_b$, where $m_b$ is the running
bottom mass at the scale $M_t$.

The two CP-even and the charged Higgs masses
read

\begin{eqnarray}
m^2_{h(H)} &=& \frac{Tr M^2 \mp \sqrt{(TrM^2)^2 - 4 \det M^2}}{2}
\label{mhH}
\end{eqnarray}
\begin{eqnarray}
 m^2_{H^{\pm}} &=& m_A^2 + (\lambda_5 - \lambda_4) v^2,
\label{mhch}
\end{eqnarray}
where
\begin{equation}
         TrM^2 =  M_{11}^2 + M_{22}^2 \;\; ; \;\;\;\;\;
     \det M^2 = M_{11}^2 M_{22}^2 - \left( M_{12}^2 \right)^2,
\label{detm2}
\end{equation}
with
 \begin{eqnarray}
  M^2_{12} &=&  2 v^2 [\sin \beta \cos \beta (\lambda_3 + \lambda_4) +
     \lambda_6 \cos^2 \beta + \lambda_7 \sin^2 \beta ] -
     m_A^2 \sin \beta \cos \beta
 \nonumber\\
     M^2_{11} &=& 2 v^2 [\lambda_1  \cos^2 \beta + 2
     \lambda_6  \cos \beta \sin \beta
     + \lambda_5 \sin^2 \beta] + m_A^2 \sin^2 \beta \\
     M^2_{22} &=&  2 v^2 [\lambda_2 \sin^2 \beta +2 \lambda_7  \cos \beta
     \sin \beta + \lambda_5 \cos^2 \beta] + m_A^2 \cos^2 \beta . \nonumber
 \end{eqnarray}
The mixing angle $\alpha$ is equally determined by
\begin{eqnarray}
      \sin 2\alpha = \frac{2M_{12}^2}
{\sqrt{\left(Tr M^2\right)^2-4\det M^2}}
 \end{eqnarray}
\begin{eqnarray}
      \cos 2\alpha = \frac{M_{11}^2-M_{22}^2}
{\sqrt{\left(Tr M^2\right)^2-4\det M^2}}
\end{eqnarray}

The above quartic couplings are given by
\begin{eqnarray}
  \lambda_1 &=& \frac{g_1^2 + g_2^2}{4} \left(1-\frac{3}{8 \pi^2} \;h_b^2 \; t
                 \right)
  \nonumber\\
        &+ &  \frac{3}{8 \pi^2}\; h_b^4\; \left[
         t + \frac{X_{b}}{2} + \frac{1}{16 \pi^2}
        \left( \frac{3}{2} \;h_b^2 + \frac{1}{2}\;h_t^2
     - 8\; g_3^2 \right) \left( X_{b}\;t + t^2\right) \right]
    \nonumber\\
  &- &   \frac{3}{96\pi^2} \; h_t^4\;\frac{\mu^4}{M_{\rm SUSY}^4}
        \left[ 1+ \frac{1}{16 \pi^2} \left( 9\;h_t^2 -5  h_b^2
     -  16 g_3^2 \right) t  \right]
\label{lambda1}
\end{eqnarray}

\begin{eqnarray}
      \lambda_2 &=& \frac{g_1^2 + g_2^2}{4} \left(1-
\frac{3}{8 \pi^2} \;h_t^2
       \; t\right)
  \nonumber\\
        &+& \frac{3}{8 \pi^2}\; h_t^4\; \left[
         t + \frac{X_{t}}{2} + \frac{1}{16 \pi^2}
        \left( \frac{3 \;h_t^2}{2} + \frac{h_b^2}{2}
     - 8\; g_3^2 \right) \left( X_{t}\;t + t^2\right) \right]
    \nonumber\\
  &- &  \frac{3}{96\pi^2} \; h_b^4\;\frac{\mu^4}{M_{\rm SUSY}^4}
        \left[ 1+ \frac{1}{16 \pi^2} \left(9\;h_b^2 -5  h_t^2
     -  16 g_3^2 \right) t  \right]
\label{lambda2}
\end{eqnarray}

\begin{eqnarray}
    \lambda_3 &=& \frac{g_2^2 - g_1^2}{4} \left(1-
\frac{3}{16 \pi^2}(h_t^2 + h_b^2) \; t \right)
\nonumber\\
        &+& \frac{6}{16 \pi^2}\; h_t^2\; h_b^2\; \left[
         t + \frac{A_{tb}}{2} + \frac{1}{16 \pi^2}
        \left( h_t^2 + h_b^2
    - 8\; g_3^2 \right) \left( A_{tb}\;t + t^2\right) \right]
  \nonumber\\
  &+ &\frac{3}{96\pi^2} \; h_t^4\; \left[\frac{3 \mu^2}{M_{\rm SUSY}^2}
                - \frac{\mu^2 A_t^2}{M_{\rm SUSY}^4} \right]
        \left[ 1+ \frac{1}{16 \pi^2} \left (6\;h_t^2 -2  h_b^2
     -  16 g_3^2 \right) t  \right]
 \nonumber\\
  &+ & \frac{3}{96\pi^2} \; h_b^4\; \left[\frac{3 \mu^2}{M_{\rm SUSY}^2}
              - \frac{\mu^2 A_b^2}{M_{\rm SUSY}^4} \right]
        \left[ 1+ \frac{1}{16 \pi^2} \left (6\;h_b^2 -2  h_t^2
     -  16 g_3^2 \right) t  \right]
\label{lambda3}
\end{eqnarray}

\begin{eqnarray}
       \lambda_4 &=&  -\; \frac{g_2^2}{2} \left(1
-\frac{3}{16 \pi^2}
       (h_t^2 + h_b^2) \; t\right)
\nonumber\\
        &-& \frac{6}{16 \pi^2}\; h_t^2\; h_b^2\; \left[
         t + \frac{A_{tb}}{2} + \frac{1}{16 \pi^2}
        \left( h_t^2 + h_b^2
     - 8\; g_3^2 \right) \left( A_{tb}\;t + t^2\right) \right]
   \nonumber\\
  &+ &\frac{3}{96\pi^2} \; h_t^4\; \left[\frac{3 \mu^2}{M_{\rm SUSY}^2}
                         - \frac{\mu^2 A_t^2}{M_{\rm SUSY}^4} \right]
        \left[ 1+ \frac{1}{16 \pi^2} \left (6\;h_t^2 -2  h_b^2
     -  16 g_3^2 \right) t  \right]
 \nonumber\\
  &+ &\frac{3}{96\pi^2} \; h_b^4\; \left[\frac{3 \mu^2}{M_{\rm SUSY}^2}
                         - \frac{\mu^2 A_b^2}{M_{\rm SUSY}^4} \right]
        \left[ 1+ \frac{1}{16 \pi^2} \left (6\;h_b^2 -2  h_t^2
     -  16 g_3^2 \right) t  \right]
\label{lambda4}
\end{eqnarray}

\begin{eqnarray}
 \lambda_5 &=& -\; \frac{3}{96\pi^2} \; h_t^4\;
                   \frac{\mu^2 A_t^2}{M_{\rm SUSY}^4}
        \left[ 1- \frac{1}{16 \pi^2} \left (2  h_b^2 - 6\;h_t^2
     +  16 g_3^2 \right) t  \right]
 \nonumber\\
  &- &\frac{3}{96\pi^2} \; h_b^4 \;
                         \frac{\mu^2 A_b^2}{M_{\rm SUSY}^4}
        \left[ 1- \frac{1}{16 \pi^2} \left ( 2  h_t^2 -6\;h_b^2
     +  16 g_3^2 \right) t  \right]
\label{lambda5}
\end{eqnarray}

\begin{eqnarray}
         \lambda_6 &=& \frac{3}{96 \pi^2}\; h_t^4\;
       \frac{\mu^3 A_t}{M_{\rm SUSY}^4}
   \left[1- \frac{1}{16\pi^2}
   \left(\frac{7}{2} h_b^2  -\frac{15}{2} h_t^2 + 16 g_3^2 \right)
        t \right]
\label{lambda6}
\\  &+ &
\frac{3}{96 \pi^2}\; h_b^4\;
\frac{\mu}{M_{\rm SUSY}} \left(\frac{A_b^3}{M_{\rm SUSY}^3}
                                            - \frac{6 A_b}{M_{\rm SUSY}}\right)
   \left[1- \frac{1}{16\pi^2}
\left(\frac{1}{2} h_t^2  -\frac{9}{2} h_b^2 + 16 g_3^2 \right)
       t  \right]
\nonumber
\end{eqnarray}

\begin{eqnarray}
\lambda_7 &=& \frac{3}{96 \pi^2}\; h_b^4\;\frac{\mu^3 A_b}{M_{\rm SUSY}^4}
   \left[1- \frac{1}{16\pi^2}
\left(\frac{7}{2} h_t^2  -\frac{15}{2} h_b^2 + 16 g_3^2 \right)
       t  \right]
\label{lambda7}
\\ &+ &
\frac{3}{96 \pi^2}\; h_t^4\;\frac{\mu}{M_{\rm SUSY}}
\left(\frac{A_t^3}{M_{\rm SUSY}^3}
         - \frac{6 A_t}{M_{\rm SUSY}}\right)
   \left[1- \frac{1}{16\pi^2}
\left(\frac{1}{2} h_b^2  -\frac{9}{2} h_t^2 + 16 g_3^2 \right)
       t  \right],
\nonumber
\end{eqnarray}
which contain the same kind of corrections as Eq.~(\ref{mhsm}),
including the leading $D$-term contributions, and where
we have defined,

\begin{eqnarray}
X_{t} & = & \frac{2 A_t^2}{M_{\rm SUSY}^2}
                  \left(1 - \frac{A_t^2}{12 M_{\rm SUSY}^2} \right) \\
          X_{b}& =& \frac{2 A_b^2}{M_{\rm SUSY}^2}
                  \left(1 - \frac{A_b^2}{12 M_{\rm SUSY}^2} \right)
\nonumber\\
      A_{tb}& =& \frac{1}{6} \left[-\frac{ 6 \mu^2}{M_{\rm SUSY}^2}
        - \frac{(\mu^2 - A_b A_t)^2}{ M_{\rm SUSY}^4}
     + \frac{3 (A_t + A_b)^2}{M_{\rm SUSY}^2}\right]. \nonumber
\end{eqnarray}

All quantities in the approximate formulae are defined at the scale
$M_t$. In particular $\alpha_3(M_t)$ is given in Eq.~(\ref{alfa3sm}),
and
\begin{eqnarray}
\label{yukawas}
h_t & = & m_t(M_t)/ (v \sin \beta)\nonumber \\
h_b & =&  m_b(M_t)/ (v \cos \beta)
\end{eqnarray}
are the top and bottom Yukawa couplings in the two-Higgs doublet model.

For $m_A\leq M_t$,  $\tan\beta$ is fixed at the scale $m_A$,
while for $m_A\geq M_t$,  $\tan\beta$ is given by~\cite{B}
\be
\label{tanbeta}
\tan\beta(M_t)=\tan\beta(m_A)\left[1+\frac{3}{32\pi^2}(h_t^2 - h_b^2)
\log\frac{m_A^2}{M_t^2}\right].
\ee

For the case in which the CP-odd Higgs mass $m_A$ is lower than
$M_{\rm SUSY}$, but still  larger than the top-quark mass scale,
we decouple, in the numerical computations, the heavy Higgs doublet
and define an effective quartic coupling for the light Higgs,
which is related to the running  Higgs mass at the scale $m_A$
through
\begin{equation}
\lambda(m_A) = \frac{m_h(m_A)}{2 v^2}.
\end{equation}
The low energy value of the quartic coupling is then obtained by
running the SM renormalization-group equations from the scale
$m_A$ down to the scale $M_t$. In the analytical approximation,
for simplicity we ignore the effect of decoupling the
heavy Higgs doublet at an intermediate scale. We partially
compensate the effect of this approximation by relating
the value of $\tan\beta$ at the scale $M_t$ with its
corresponding value at the scale $m_A$ through its renormalization-group
running, Eq.~(\ref{tanbeta}).

In Figs.~3, 4 and 5 we  plot $m_h$, $m_H$ and $m_H^{\pm}$,
respectively, as functions of $m_A$, for $M_t=175$ GeV and the four
cases corresponding to $\tan\beta=15$ and $\tan\beta \simeq$ 1.6
(which is the lowest possible value of $\tan \beta$ for this $M_t$ and
corresponds to the IR fixed point solution) and for large and zero mixing.
Solid curves correspond to the numerical
renormalization-group improved  LL approximation and dashed
curves to the analytical approximation depicted in the previous
formulae. For the charged Higgs mass, we have also plotted the
case $\mu = 2 M_{\rm SUSY}$ and $\tan\beta = 15$,  since, unless
large values of $\tan\beta \simeq m_t/m_b$ are considered,
its  radiative
corrections depend on $\mu$ but not on $A_t$ or $A_b$
(see Eqs.~(\ref{lambda4}) and (\ref{lambda5})).
We can observe that, in all cases, the agreement between the
Higgs masses in the numerical renormalization-group improved calculation
and the analytical approximation is, either of the order of,
or better than,  3~GeV.

\vskip  0.5 cm
\begin{center}
\begin{tabular}{||c|c||}\hline
Vertex &  Coupling \\ \hline
  &   \\
$\{h,H\}W_{\mu}W_{\nu}$ &${\displaystyle ig_2M_Wg_{\mu\nu}
\{\sin(\beta-\alpha),
\cos(\alpha+\beta)\} }$\\\  & \\
$\{h,H\}Z_{\mu}Z_{\nu}$ & ${\displaystyle ig_2\frac{M_W}
{\cos^2\theta_W}g_{\mu\nu} \{\sin(\beta-\alpha),\cos(\alpha+\beta)\} }$\\\
 &  \\
$\{h,H,A\}u\overline{u}$ & ${\displaystyle
-\frac{i}{2}\left(\frac{m_u}{M_W}\right)\frac{g_2}{\sin\beta}
\{\cos\alpha,\sin\alpha,-i\gamma_5\cos\beta\}  }$\\\
 & \\
$\{h,H,A\}d\overline{d}$ & ${\displaystyle
-\frac{i}{2}\left(\frac{m_d}{M_W}\right)\frac{g_2}{\cos\beta}
\{-\sin\alpha,\cos\alpha,-i\gamma_5\sin\beta\}  }$\\\
& \\
$\{h,H\} A Z_{\mu}$ & ${\displaystyle -\frac{e (p + k)_{\mu}}{2 \cos\theta_W
\sin\theta_W}
\{\cos(\beta - \alpha),-\sin(\beta-\alpha)\} }$\\  &  \\
\hline
\end{tabular}
\end{center}
\vskip .5 cm

Notice that the knowledge of the radiatively corrected quartic couplings
$\lambda_i$, $i=1,\dots, 7$, and hence of the corresponding value of the Higgs
mixing
angle $\alpha$, permits the  evaluation of all radiatively corrected Higgs
couplings. For
instance some  important Yukawa and gauge Higgs couplings~\cite{R}
are listed in the table above,
where $p_{\mu}$ ($k_{\mu}$) is the incoming (outcoming) CP-odd
(CP-even) Higgs momentum. Furthermore, the trilinear Higgs
couplings can be explicitly written as functions of
$\lambda_i$, $i=1,\dots, 7$, $\alpha$ and $\beta$~\cite{HH,HHN}. To check the
goodness of our analytical approximation on the different couplings we plot
in Fig.~6, for the same four cases as in Figs.~3--5,
$\cos\alpha$   as a function of $m_A$, for $M_t=175$ GeV.
As can be seen from Fig.~6,  we find excellent agreement
between the values obtained with the numerical and analytical approximations.

\section{On the expansion variables}

In this section we shall comment on the definition of
$M_{\rm SUSY}$.
As we mentioned above, the scale $M_{\rm SUSY}$, appearing in the
quartic coupling expressions, should be associated with a characteristic
squark mass scale. The renormalization group analysis considered in
this work is based on the assumption of a single step decoupling of the
squark fields, which is valid only if the mass splitting among the
squark mass eigenstates is small. More quantitatively,
the validity of our expansion requires,
\begin{equation}
\frac{m_{\tilde{t}_1}^2 - m_{\tilde{t}_2}^2}
{m_{\tilde{t}_1}^2 + m_{\tilde{t}_2}^2} \simlt 0.5,
\label{split}
\end{equation}
where $m_{\tilde{t}_1}^2$ and $m_{\tilde{t}_2}^2$ are the
stop squared mass eigenvalues.
This restriction also applies to all existing Higgs mass analyses
at the next-to leading order  [10--12].
Observe that Eq.~(\ref{split})
automatically ensures the absence of tachyons in the stop sector.
Furthermore, under the  validity of Eq.~(\ref{split}),
the scale $M_{\rm SUSY}^2$ may be safely defined
as the arithmetic
average of the stop squared mass eigenvalues,
\begin{equation}
M_{\rm SUSY}^2 =\frac{m_{\tilde{t}_1}^2 + m_{\tilde{t}_2}^2}{2}.
\end{equation}
Observe that our work, as well as all other RG Higgs
mass analyses performed in the literature,
also relies on an expansion of the effective
Higgs potential up to operators of dimension four.  The contribution
of higher dimensional operators may be safely neglected only if
$2 |M_t A_t| < M_{\rm SUSY}^2$ and $2 |M_t \mu| < M_{\rm SUSY}^2$.

Strictly speaking, the operator expansion was performed in the symmetric
phase, where the squared of the energy scale at which the stops
decouple is not the average of the stop squared masses, $M_{\rm SUSY}$,
but the average of the stop soft supersymmetry breaking squared
mass parameters, $m^2\equiv (m_Q^2+m_U^2)/2$.
Higher dimensional operators in the
symmetric phase can only be neglected if $M_{\rm SUSY}^2 \gg M_t^2$.
In this range of parameters, the distinction between $M_{\rm SUSY}$
and the real decoupling scale has no impact on our approximations.
In general, our expressions give a reliable approximation to the
Higgs masses whenever the above restriction on the stop mixing
parameter, Eq.~(\ref{split}), holds, and $M_{\rm SUSY} \simgt 2 M_t$.
Moreover, for the large $\tan\beta$ case, for which the
bottom-Yukawa contribution
becomes relevant, we have assumed that the sbottom masses
are of order of the stop  ones and that similar bounds on their
mixing mass parameters are fulfilled.

Physically it is clear that,
in the region where higher order operators are unsuppressed,
$M_{\rm SUSY} \simlt 2 M_t$, the leading logarithm expansion
of the Higgs masses should contain
powers of $\log(M_{\rm SUSY}^2/M_t^2)$
rather than $\log(m^2/M_t^2)$.
This should be the case, since, in the supersymmetric limit
for the top/stop sector, $M_{\rm SUSY} = M_t$,
and zero mixing,
one should recover the tree level values for the Higgs masses.
One may ask how accurately our analytical expressions
may determine the Higgs masses
for values of $M_{\rm SUSY} \simlt 2 M_t$, for which,
strictly speaking, one is beyond the limits of the
present expansion. In fact,
in order to check the reliability of our approximation for the case
in which light stops, with masses of the order of the top mass, are
present, an alternative computation, valid in that regime, should
be used. The expressions for the
CP-even Higgs mass matrix elements
computed in Ref.~\cite{ERZ}, and based on the one-loop effective
potential in the MSSM (keeping the leading supersymmetric
corrections from the stop/sbottom sectors) can be useful for the
comparison.  Observe that the MSSM effective potential is valid
only at scales of the order of, or larger than, the characteristic
stop supersymmetry breaking terms.
One cannot consider  the MSSM effective potential
at scales below the squark
masses since, at those scales, the squarks are already
decoupled, possibly leaving threshold corrections to match the
effective potential beyond and below the supersymmetric threshold.

Once recognized that the MSSM effective potential is
valid at the scale defined by the  squark masses, one can try
to {\it improve} its performance by means of its RG
improvement~\cite{Ford}.
Were the RG improved effective potential exactly
scale independent, one could minimize it at any scale
(e.g. at $M_{\rm SUSY}$), defining all the running parameters on which
the effective potential depends also at that scale, and rewrite them, in turn,
as functions of the corresponding parameters defined at the
scale $M_t$. Then, taking advantage of the scale
independence of the effective potential one could run the Higgs masses from the
high scale to $M_t$ using the anomalous dimensions of the Higgs
fields. However, the RG improved one-loop effective potential is not
exactly scale independent (even after including all LL and
NTLL corrections) and hence, when minimizing it at $M_{\rm SUSY}$
(see e.g. Figs.~1 and 3 in Ref.~\cite{CEQR})
one is implicitly including an error in the lightest CP-even Higgs
mass determination. This error is very small for values of $M_{\rm SUSY}$
of the order of the weak scale, but becomes significant for larger scales
(e.g. $M_{\rm SUSY}\simgt 500$ GeV).

We therefore consider the RG improved MSSM effective potential,
which is minimized at the scale $M_{\rm SUSY}$, at which we determine
the neutral Higgs boson squared mass matrix. All the parameters in the
mass matrix are running parameters, defined at the scale $M_{\rm SUSY}$
(including the vacuum expectation values $v_1$ and $v_2$).
They are related to parameters defined at the scale $M_t$ by the
corresponding evolution with the $\beta$- and $\gamma$-functions, that we
compute to one-loop LL (i.e. first order in the expansion parameter
$\log(M_{\rm SUSY}^2/M_t^2)$) and keeping only the Yukawa and strong
gauge couplings, as in our analytical expressions of sections 2 and 3.
We then evolve the squared mass matrix, with the
corresponding anomalous dimension matrix,
from $M_{\rm SUSY}$ to $M_t$.
In Fig.~7 we plot
the lightest CP-even Higgs mass obtained from our
analytical expressions (solid lines) and that obtained from the
RG improved effective potential of the MSSM (dashed lines),
for a pole top-quark mass
$M_t=175$ GeV, vanishing mixing $A_t=\mu=0$,
$m_A=M_{\rm SUSY}$ (to simplify
the comparison), and large ($\tan\beta=15$)
and small (IR) values of $\tan\beta$.
We can see that
the agreement between the Higgs masses obtained by using our
analytical expressions and the RG improved MSSM effective potential
is very good for all values of $M_{\rm SUSY}\simlt 500$ GeV,
and in particular for
$M_{\rm SUSY} \simlt 2 M_t$,
for which the latter should give
a reliable description. Hence, we see that, independently of
the value of $\tan\beta$, our analytical expressions give a
very good approximation to the Higgs mass values,
in the case of no left-right stop mixing,
even in the regime $M_{\rm SUSY}\simlt 2 M_t$.
For the case of stop mixing considered in Fig.~8,
observe that, independently of the value of $\tan\beta$,
the results are also in very good
agreement for values of $M_{\rm SUSY} \simlt
500$ GeV. Similar conclusions may be obtained for any value
of the stop mixing parameter.

Observe that, in order to compute pole Higgs masses, Higgs vacuum
polarization contributions should be added to our computation.
As we discussed in section 2, by fixing the renormalization scale to
the pole top-quark mass, we make the top-dependent vacuum
polarization contributions small. The stop-dependent
contributions are renormalization-scale independent.
For light stops, these contributions
are opposite in sign to the top leading ones and of
the same order of magnitude; they become very small
if the stops are heavy. In general, for our choice of the
renormalization scale, the vacuum polarization contributions are
small, and would only be required in a computation of
Higgs masses going beyond the approximation presented in the
present work.

Summarizing, even for stop masses
of order of the top mass, our analytical expressions give a very
good approximation to the Higgs mass value. Observe that, since
we have applied one step decoupling on the stops,
the constraint on the stop mass splitting, Eq.~(\ref{split}), should
always be held.
Finally, in Figs.~7 and 8 we also plot, for comparison,
the values obtained from
the MSSM effective potential while ignoring the RG improvement,
and evaluating all running masses and couplings directly at the scale $M_t$
(dotted lines). In particular, we use, for the top mass, the running value
$m_t(M_t)$. This approximation overestimates the Higgs mass
for basically all values of $M_{\rm SUSY}$.

\section{Conclusions}

In this work we present a one-loop improved
renormalization-group analysis of the lightest CP-even Higgs mass,
which, for the case $m_A \sim M_{\rm SUSY}$,
reproduces the existing
two-loop renormalization-group improved effective potential
results within an error of less than 2~GeV.
For the general case
$m_A\simlt M_{\rm SUSY}$, we  provide an analytical
approximation (including two-loop leading-log corrections)
for the effective quartic couplings, which determines the Higgs
mass spectrum within an error of less than 3 GeV with respect to the
values obtained  using the improved
one loop RG approximation.
Our formulae are also useful to write
radiatively corrected expressions for the Higgs couplings, which
are of interest in computing physical processes.

Our analysis takes into account the
dominant top and bottom quark Yukawa
(as well as strong gauge) coupling effects, by assuming
a characteristic supersymmetric scale, $M_{\rm SUSY}$,
and including squark mixing effects.
In some realistic scenarios the chargino
and neutralino masses may become much smaller than the
characteristic squark mass scale. Light charginos
and neutralinos have a small and negative effect on the
lightest CP-even Higgs mass, of order 2--3 GeV~\cite{ERZ,Marc},
which may be
added to our analytical approximation at the one-loop order.

Our approximation relies on a one step decoupling of all
supersymmetric particles and, in particular, on
the stop mixing parameter requirements discussed in section 4.
A computation of the Higgs masses and couplings going beyond
the approximation presented in this work, would demand the derivation
of the full two-loop RG improvement of the one-loop effective potential,
and polarization corrections to running Higgs masses,
including the squark and heavy Higgs contributions, the analogous
to what has been done for the case of large squark and CP-odd Higgs
masses in Ref.~\cite{CEQR}.
We expect to come back to this subject
elsewhere.

\section*{Acknowledgements}
The work of J.R.E. is supported by the Alexander-von-Humboldt Stiftung.
The work of C.W. and M.C. is partially supported by the Worldlab.
We wish to thank J.-F.~Grivaz, H.E.~Haber, R.~Hempfling,  A.~Hoang
and F.~Zwirner for useful discussions  and comments.

%%%%%%%%%%%%%%%%%%%%%%%%%%%%%%%%%%%%%%%%%%%%%%%%%%%
\begin{figure}
\centerline{
\psfig{figure=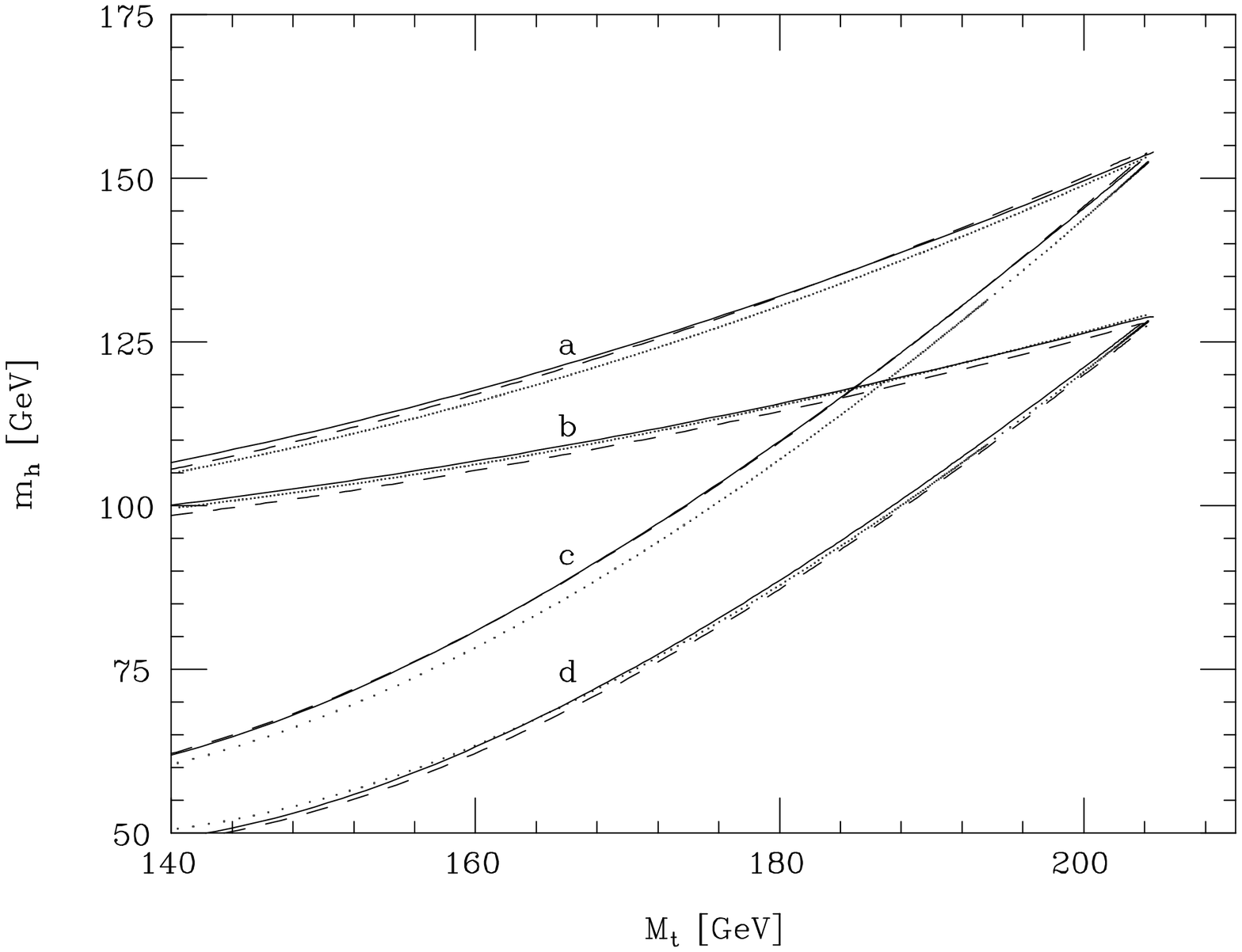,width=20cm,height=15cm,angle=90} }
\caption[0]
{The lightest CP-even Higgs mass as  a function of the physical top-quark
mass, for $M_{\rm SUSY}=1$ TeV,
evaluated in the limit of large CP-odd Higgs mass, as obtained
from the two-loop renormalization-group improved
effective potential (solid lines), the one-loop improved RG evolution
(dashed lines) and the analytical formulae, Eq.~(\ref{mhsm}) (dotted lines).
The four sets of lines correspond to:
{\bf a)} $\tan \beta =$ 15 with maximal squark mixing, $\tilde{X}_t$ = 6;
{\bf b)} $\tan \beta =$ 15 with zero mixing, $\tilde{X}_t$ = 0;
{\bf c)} the minimal value of $\tan \beta$ allowed by
perturbativity constraints for the given value of $M_t$ (IR fixed point),
with $\tilde{X}_t$ = 6; and,
{\bf d)} $\tan \beta $ the same as in (c) with  zero mixing.}
\end{figure}

\begin{figure}
\centerline{
\psfig{figure=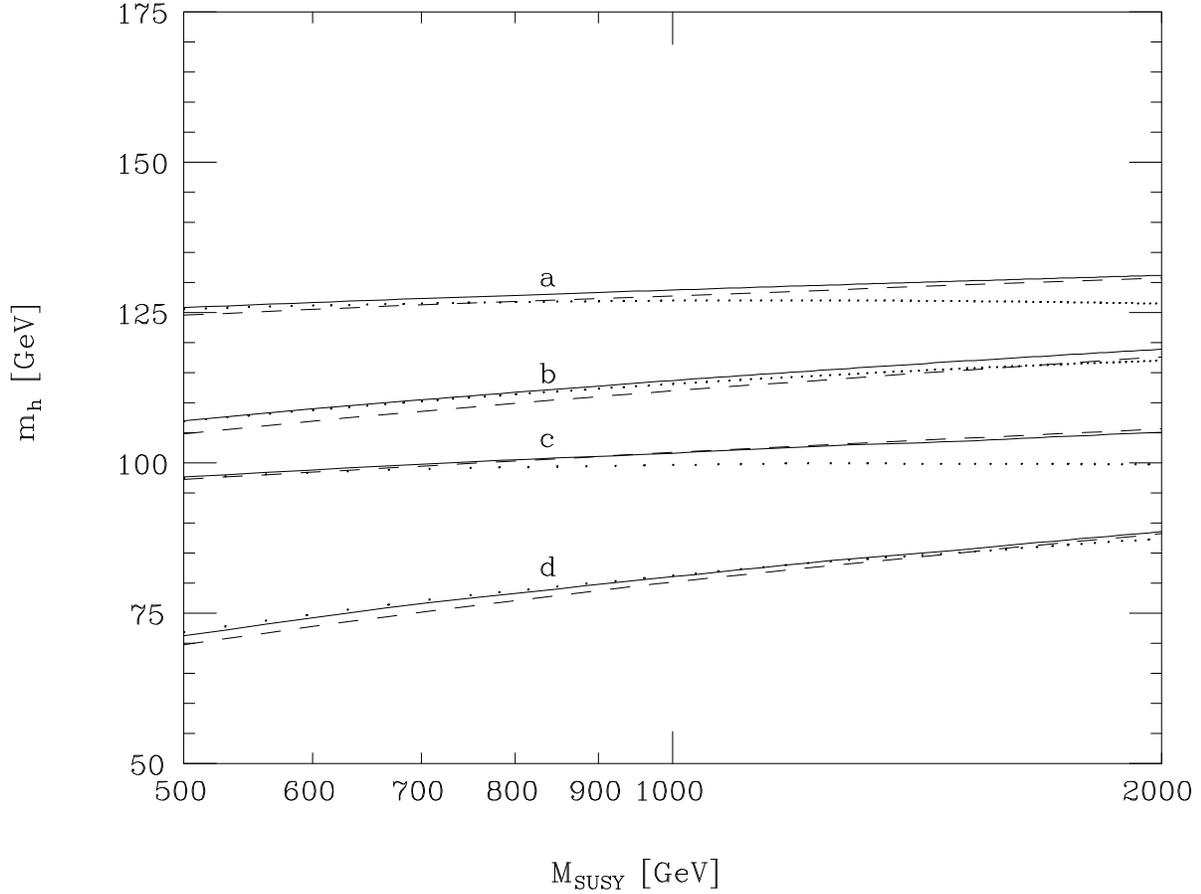,width=20cm,height=15cm,angle=90} }
\caption[0]
{The lightest CP-even Higgs mass as  a function of the supersymmetric scale
$M_{\rm SUSY}$, for $M_t=175$ GeV,
and evaluated in the limit of large CP-odd Higgs mass.
Solid, dashed and dotted lines are as in Fig.~1. The different
sets of curves correspond to the values of $\tilde{X}_t$ and
$\tan\beta$ for cases (a) to (d) of Fig.~1.}
\end{figure}

\begin{figure}
\centerline{
\psfig{figure=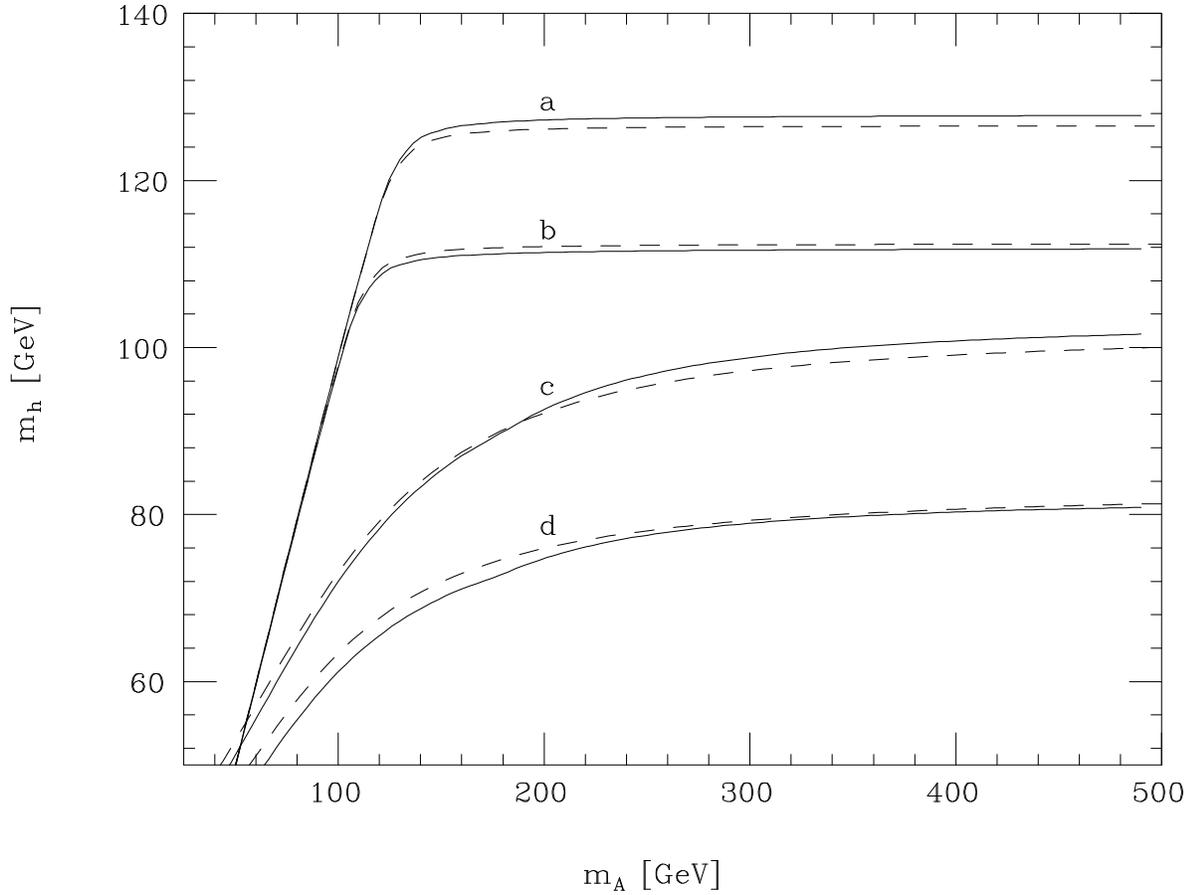,width=20cm,height=15cm,angle=90} }
\caption[0]
{The lightest CP-even Higgs mass as  a function of the
 CP-odd Higgs mass for
 a physical top-quark mass $M_t =$ 175 GeV and $M_{\rm SUSY}$ = 1 TeV, as
obtained from the one-loop improved RG evolution
(solid lines) and the analytical formulae, Eq.~(\ref{mhH}), (dashed lines).
The four sets of curves correspond to:
{\bf a)} $\tan \beta=$ 15 with large squark mixing, $X_t$ = 6
($\mu=0$);
{\bf b)} $\tan \beta=$ 15 with zero mixing $X_t=\mu$ = 0;
{\bf c)}
the minimal value of $\tan \beta$ allowed by
perturbativity constraints for the given value of $M_t$ (IR fixed point),
$\tan\beta\simeq 1.6$, with large squark mixing; and,
{\bf d)} $\tan\beta\simeq 1.6$ with zero mixing.  }
\end{figure}

\begin{figure}
\centerline{
\psfig{figure=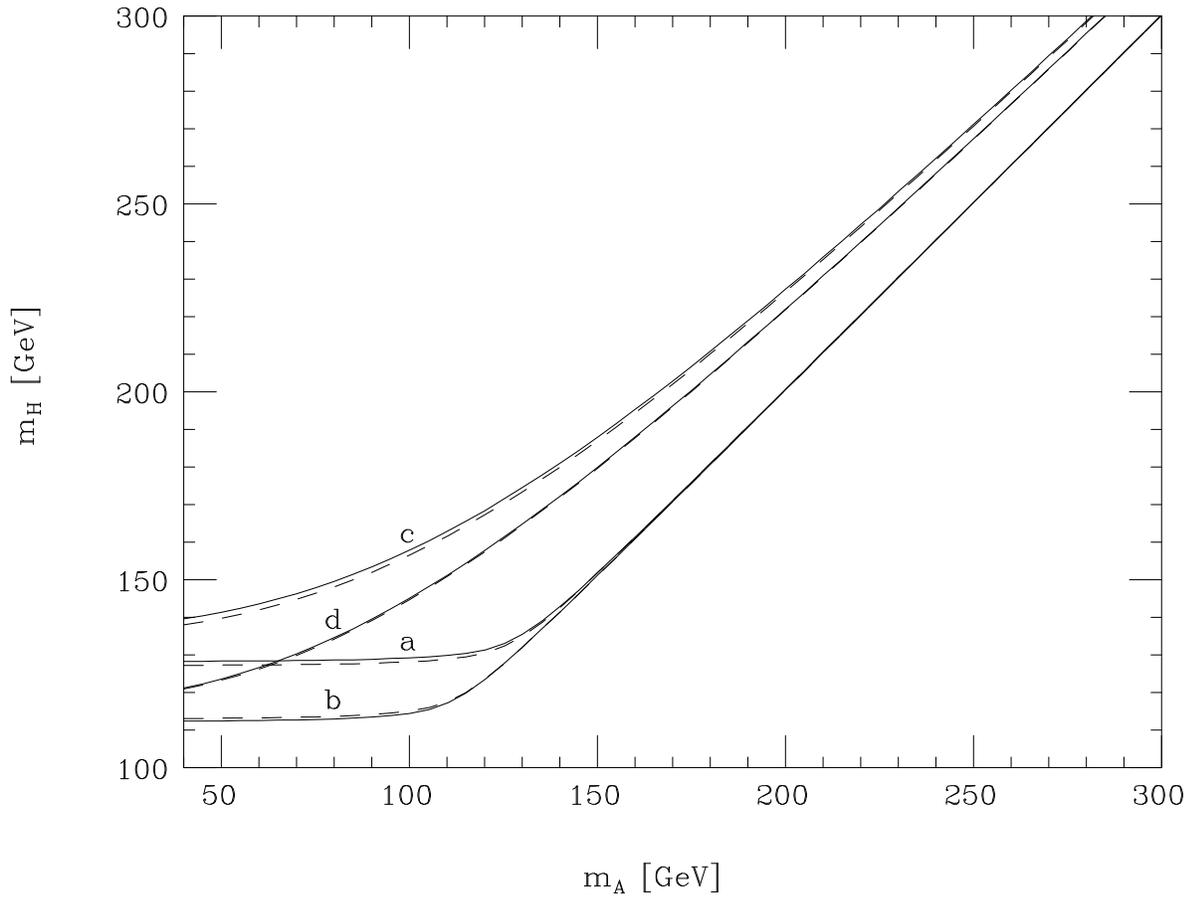,width=20cm,height=15cm,angle=90} }
\caption[0]{The heaviest CP-even Higgs mass as  a function of the
CP-odd Higgs mass for the same set of parameters
as in Fig.~3.}
\end{figure}

\begin{figure}
\centerline{
\psfig{figure=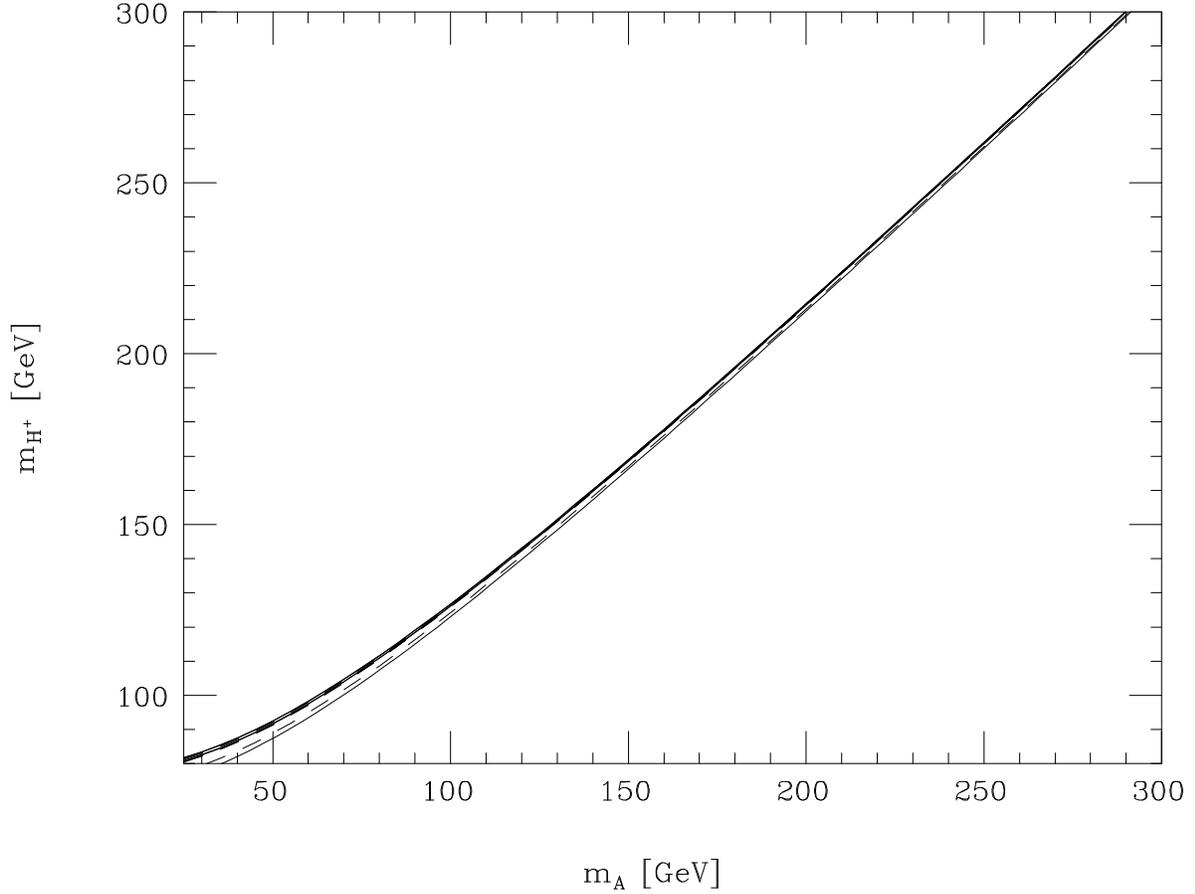,width=20cm,height=15cm,angle=90} }
\caption[0]{The charged Higgs mass as  a function of the
CP-odd Higgs mass for the same set of parameters
as in Fig.~3
(upper overlapping curves), and for
$\mu = 2 M_{\rm SUSY}$ and $\tan\beta = 15$, for which the radiative
corrections become observable (solid and dashed lower curves).}
\end{figure}

\begin{figure}
\centerline{
\psfig{figure=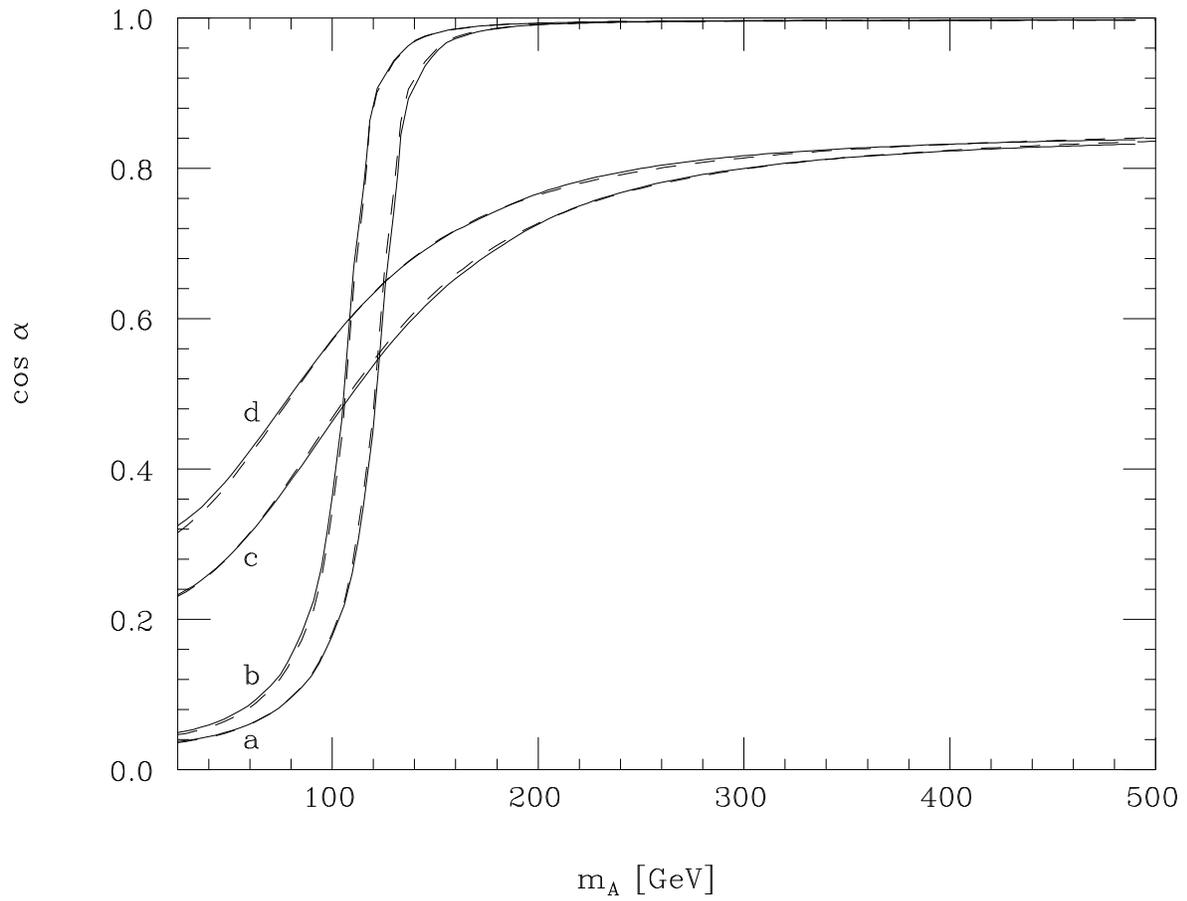,width=20cm,height=15cm,angle=90} }
\caption[0]{Plot of $\cos\alpha$, as a function
of the CP-odd Higgs mass for the same set of parameters as in Fig.~3.}
\end{figure}

\begin{figure}
\centerline{
\psfig{figure=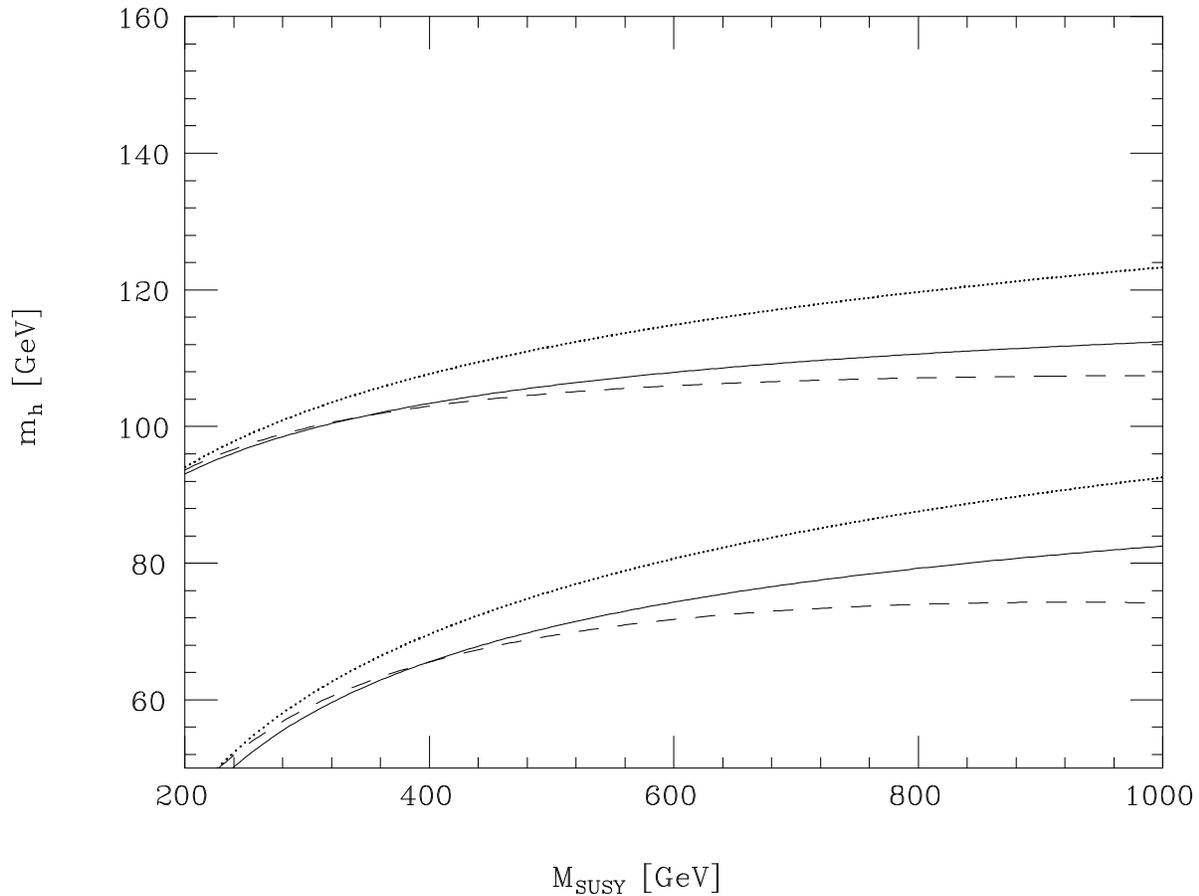,width=20cm,height=15cm,angle=90}}
\caption[0]{Comparison between the results for the Higgs mass
obtained with the analytical approximation (solid lines), the
RG improved one-loop MSSM effective
potential (dashed lines) and the unimproved one-loop MSSM
effective potential (dotted lines),
as described in section 4,
for $M_t=175$ GeV,
$m_A = M_{\rm SUSY}$, $m_Q=m_U$,
and $A_t = \mu =0$. The lower set
corresponds to
$\tan\beta=1.6$, and the upper set to $\tan\beta = 15$.}
\end{figure}

\begin{figure}
\centerline{
\psfig{figure=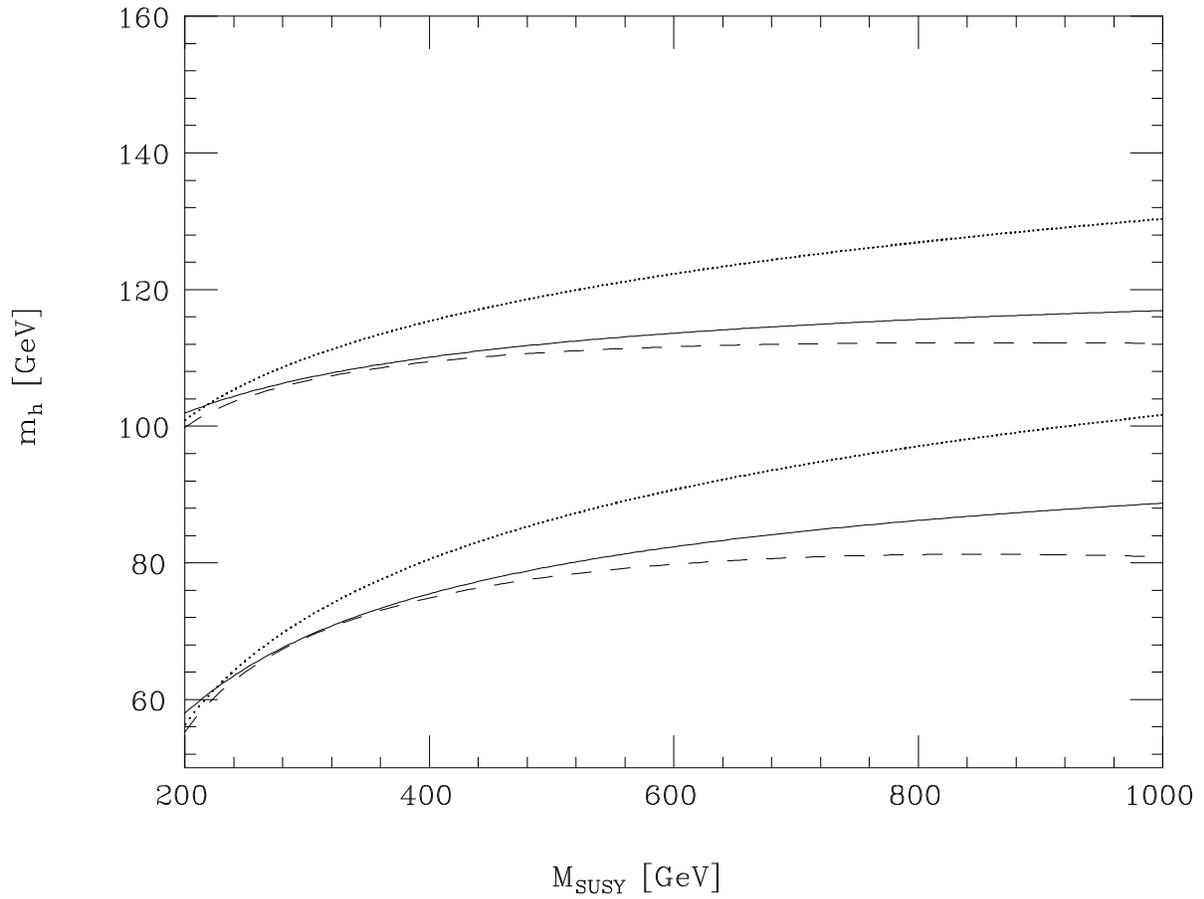,width=20cm,height=15cm,angle=90}}
\caption[0]{The same as in Fig. 7 but for $A_t = M_{\rm SUSY}$.}
\end{figure}

\end{document}